\documentclass[aps,prd,onecolumn,groupedaddress,floatfix,longbibliography,superscriptaddress,notitlepage]{revtex4-2}

\usepackage{amsfonts}
\usepackage{amsmath}
\usepackage{amssymb}
\usepackage{amsfonts}
\usepackage{graphicx}
\usepackage{bbold}
\usepackage{bm}
\usepackage{cancel}
\usepackage{times,float}
\usepackage{graphicx}
\graphicspath{{FIGURES/}}
\usepackage[dvipsnames,svgnames]{xcolor}
\usepackage{hyperref}
\hypersetup{colorlinks=true, linkcolor=Blue, citecolor=Blue,urlcolor=Blue}
\usepackage{multirow}
\usepackage{ulem}
\usepackage{color,epsfig}
\usepackage{bm}
\usepackage{slashed}
\usepackage{hyperref}
\usepackage{feynmf}
\usepackage{array}
\usepackage{physics}
\usepackage{subcaption}
\allowdisplaybreaks
\usepackage{comment}
\usepackage[none]{hyphenat}
\usepackage{hyperref}
\hypersetup{colorlinks=true, linkcolor=Blue, citecolor=Blue,urlcolor=Blue}

\begin{document}

\title{Fermionic Casimir effect in an axial Lorentz-violating background}

\author{A. Mart\'{i}n-Ruiz}
\email{alberto.martin@nucleares.unam.mx}
\affiliation{Instituto de Ciencias Nucleares, Universidad Nacional Aut\'{o}noma de M\'{e}xico, 04510 Ciudad de M\'{e}xico, M\'{e}xico}

\author{M. B. Cruz}
\email{messiasdebritocruz@servidor.uepb.edu.br}
\affiliation{Universidade Estadual da Paraíba (UEPB), Centro de Ciências Exatas e Sociais Aplicadas (CCEA), R. Alfredo Lustosa Cabral, s/n, Salgadinho, Patos - PB, 58706-550 - Brazil.}

\author{E. R. Bezerra de Mello}
\email{emello@fisica.ufpb.br}
\affiliation{Departamento de Física, Universidade Federal da Paraíba,
Caixa Postal 5008, Jo\~ao Pessoa, Paraíba, Brazil}

\begin{abstract}
We investigate the fermionic Casimir effect for a Dirac field confined between two parallel plates with MIT bag boundary conditions in the presence of CPT-odd Lorentz-symmetry violation described by a constant axial background vector $b_{\mu}$. The exact mode quantization is derived from the modified Dirac equation in the planar geometry, and the vacuum energy is formulated through a phase-shift representation. For spacelike backgrounds we show that the components parallel to the plates can be absorbed into a shift of the transverse momenta and therefore do not affect the renormalized Casimir energy, while the component normal to the plates modifies the longitudinal spectrum and produces a genuine Lorentz-violating correction. Both the timelike component $b_{0}$ and the normal spacelike component $b_{z}$ can thus be treated within a unified framework characterized by a single effective spectral parameter. A closed logarithmic integral representation for the Casimir energy is obtained and its behavior is analyzed in the Lorentz-symmetric, weak-background, and strong-background regimes.
\end{abstract}

\maketitle

%--------------- Introduction --------------
\section{Introduction}

Quantum fluctuations of the vacuum are modified whenever a quantum field is confined by external boundaries. A paradigmatic manifestation of this phenomenon is the Casimir effect, originally predicted by Casimir for the electromagnetic field between two parallel conducting plates \cite{Casimir1948}. Early experimental confirmations were limited in precision \cite{Sparnaay1958}, but subsequent high-accuracy measurements \cite{Lamoreaux:1996wh,Mohideen:1998iz,Bressi:2002fr,Decca:2003zk} firmly established the effect and turned it into a benchmark phenomenon in quantum field theory (QFT) and fluctuation-induced interactions. In general, the presence of boundaries alters the spectrum of normal modes of a quantum field and leads to observable stresses on the confining surfaces. Over the past decades, the Casimir effect has been extensively investigated for different geometries, boundary conditions, and types of quantum fields, as well as in nontrivial backgrounds \cite{Milton2001,Bordag2009,Plunien:1986ca,Klimchitskaya:2009zz,Birrell:1982ix}.

Because the Casimir energy is determined by the spectrum of vacuum fluctuations, it is particularly sensitive to modifications of the underlying field dynamics. For this reason, Casimir systems provide a useful theoretical laboratory for probing extensions of relativistic QFT. In particular, considerable attention has been devoted to studying the Casimir effect in frameworks where Lorentz symmetry is violated. Lorentz invariance plays a fundamental role in both the Standard Model of particle physics and general relativity, yet several candidate theories of quantum gravity suggest that it may be violated at very high energies. A systematic way to describe possible deviations from Lorentz invariance is provided by the Standard-Model Extension (SME), which incorporates Lorentz-violating operators in an effective-field-theory framework \cite{Colladay1997,Colladay1998,Kostelecky2004}. In addition, Lorentz symmetry breaking can arise in string-inspired scenarios and in quantum-gravity–motivated models with modified dispersion relations or preferred frames \cite{Kostelecky:1988zi,Jacobson:2000xp,Mattingly:2005re,Liberati:2013xla}.

In Lorentz-violating theories, the presence of fixed background tensors modifies the propagation of quantum fields by introducing preferred directions in spacetime. These modifications typically manifest through anisotropic dispersion relations or through couplings that shift the energy spectrum of the field modes. Since the Casimir effect depends directly on the mode structure of the vacuum, it is naturally affected by such background structures. A number of works have therefore explored how Lorentz-violating operators modify Casimir energies and forces in scalar \cite{PhysRevD.96.045019,Erdas:2020ilo,Escobar2020, ESCOBAR2020135567}, fermionic \cite{Frank2008, 10.1093/ptep/ptae016, PhysRevD.99.085012} and electromagnetic theories \cite{Kharlanov2009,Escobar2020,PhysRevD.94.076010} and in a scenario with a bumblebee field coupled to gravity \cite{Tang:2025eew}. In this context, important contributions regarding the behavior of quantum fields under boundaries and axial backgrounds have also been developed in Refs.~\cite{Dudal2018, Dudal:2020yah, Dudal2024}, which enrich the understanding of topological and boundary degrees of freedom.

Interestingly, field theories containing Lorentz-violating background vectors are not only relevant in high-energy physics but also arise naturally in condensed-matter systems. In particular, the low-energy quasiparticles in Dirac and Weyl materials are described by relativistic-like Hamiltonians in which symmetry-breaking parameters play the role of effective background fields. For example, in Weyl semimetals the separation of Weyl nodes in momentum or energy space, as well as the breaking of inversion or time-reversal symmetry that distinguishes the two chiral sectors, can be encoded through axial-vector terms formally analogous to those appearing in Lorentz-violating extensions of relativistic fermion theories \cite{Armitage2018,Grushin2012,Goswami2013,Hasan2010,Qi2011}. This correspondence has stimulated growing interest in the interplay between relativistic field-theory phenomena and emergent quasiparticle dynamics in topological materials \cite{PhysRevResearch.4.023106,PhysRevD.109.065005,sym17040581}. In this condensed-matter context, the physical interfaces of a thin-film or slab of a Weyl semimetal naturally implement the MIT bag boundary conditions, as the abrupt termination of the crystal lattice prevents the quasiparticles from escaping into the vacuum, thereby strictly vanishing the normal fermionic current \cite{ElizaldeKirsten}. The resulting size-dependent vacuum energy corrections can be probed through modern high-precision experimental techniques, such as monitoring the resonant frequency shifts in nano-electromechanical systems (NEMS) \cite{Capasso:2007zz} or tracking thermodynamic anomalies via magnetic torque spectroscopy under strain-modulated conditions \cite{Woods:2016vwu}.

In this work we analyze the fermionic Casimir effect for a Dirac field in the presence of a CPT-odd axial background vector $b_{\mu}$. Such a term appears in the fermion sector of the SME and modifies the Dirac equation through an axial coupling that alters the dispersion relations of the fermionic modes. We consider a planar geometry in which the fermionic field is confined between two parallel plates satisfying MIT bag boundary conditions. By solving the modified Dirac equation in this geometry we obtain the exact quantization condition for the longitudinal momentum and construct the corresponding vacuum energy.

Our analysis shows that the physical impact of the background depends strongly on its orientation relative to the plates. For spacelike backgrounds, components parallel to the plates can be absorbed into a shift of the transverse momenta and therefore do not contribute to the Casimir energy. In contrast, the component normal to the plates modifies the longitudinal spectrum and produces a genuine Lorentz-violating correction. We show that both the timelike component $b_0$ and the normal spacelike component $b_z$ can be treated within a unified spectral framework characterized by a single effective parameter. To evaluate the vacuum energy we employ a density-of-states formulation based on the phase shift associated with the modified quantization condition. This approach allows us to derive a closed logarithmic representation of the Casimir energy and to analyze its behavior in different regimes, including the Lorentz-symmetric limit as well as weak and strong background fields.

In the present work, the axial background $b_\mu$ is implemented as a constant, non-dynamical coefficient within the minimal Standard-Model Extension (SME) framework. In this effective field theory description, any dynamical fluctuations or radiative corrections coming from an underlying spontaneous Lorentz symmetry breaking mechanism are heavily suppressed on laboratory length scales, allowing us to safely treat $b_\mu$ as a rigid vector background. We note that while recent studies, such as Ref.~\cite{Muniz_2025}, have investigated Lorentz-violating vector backgrounds and Casimir-type quantum corrections in non-trivial gravitational fields , our calculation isolates the flat-space spectral modifications of the confined fermionic vacuum under explicit chiral couplings.

Compared to previous studies of the fermionic Casimir effect under Lorentz violation \cite{PhysRevD.99.085012}, the main novelty of the present work lies in the exact closed logarithmic representation of the vacuum energy density and the analytical unification of the CPT-odd timelike $b_0$ and longitudinal spacelike $b_z$ components. By implementing a phase-shift formulation tailored to the MIT bag boundaries \cite{ElizaldeKirsten}, we establish a precise geometric selection rule and derive exact asymptotic expressions that explicitly map out the suppression of vacuum fluctuations under strong axial backgrounds.

The remainder of the paper is organized as follows. In Sec.~\ref{Theory} we introduce the Lorentz-violating Dirac theory and derive the modified dispersion relations relevant for the planar geometry. In Sec.~\ref{mode_section} we determine the mode quantization conditions imposed by the MIT bag boundary conditions. In Sec.~\ref{sec:casimir_unified} we construct the vacuum energy using the density-of-states formalism and the associated phase-shift representation. We also present a closed integral form of the Casimir energy and discuss its behavior in different physical limits. Finally, Sec.~\ref{conclusions} summarizes our conclusions and outlines possible connections with both high-energy and condensed-matter realizations of Lorentz-violating fermionic systems.

%------------- Theory ---------------
\section{Axial Lorentz-violating Dirac theory} \label{Theory}

We consider a relativistic fermionic field subject to CPT-odd Lorentz symmetry violation, described by an axial background vector $b _{\mu}$. The dynamics is governed by the modified Dirac Lagrangian density \cite{Colladay1997,Colladay1998,Kostelecky2004}
\begin{align}
    \mathcal{L} = \bar{\psi} \left( i \hbar c \, \gamma ^{\mu} \partial _{\mu} - b _{\mu} \gamma ^{5} \gamma ^{\mu} \right) \psi ,  \label{Lagrangian_spacelike_axial}
\end{align}
where $\bar{\psi} = \psi ^{\dagger} \gamma ^{0}$ and the gamma matrices satisfy $\{ \gamma ^{\mu} , \gamma ^{\nu} \} = 2 \eta ^{\mu \nu}$ with $\eta ^{\mu \nu} = \mathrm{diag}(1,-1,-1,-1)$. The constant four-vector $b _{\mu}$ introduces a preferred direction in spacetime and explicitly breaks Lorentz invariance, while preserving translational invariance. This term corresponds to the CPT-odd axial coupling of the fermion sector of the Standard-Model Extension (SME).

The Euler-Lagrange equations derived from \eqref{Lagrangian_spacelike_axial} lead to the modified Dirac equation
\begin{align}
    \left( i \hbar c \, \gamma ^{\mu} \partial _{\mu} - b _{\mu}  \gamma ^{5} \gamma ^{\mu} \right) \psi (x) = 0 .    \label{Dirac_eq_spacelike_axial}
\end{align}
The axial nature of the Lorentz-violating term implies that left- and right-handed components of the Dirac field couple to $b _{\mu}$ with opposite signs, a feature that plays a central role in the spectral properties of the theory.

For axial couplings it is particularly convenient to work in the chiral (Weyl) representation of the gamma matrices,
\begin{align}
    \gamma ^{5} = \begin{pmatrix} - I & 0 \\ 0 & I \end{pmatrix} , \qquad \gamma ^{\mu} = \begin{pmatrix} 0 & \sigma ^{\mu} \\ \bar{\sigma} ^{\mu} & 0 \end{pmatrix} , \qquad \sigma ^{\mu} =( I , \boldsymbol{\sigma} ) , \quad \bar{\sigma} ^{\mu} = ( I , - \boldsymbol{\sigma} ) , \label{chiral_rep_general}
\end{align}
where $\boldsymbol{\sigma}$ represent the Pauli matrices. Decomposing the Dirac spinor as $\psi = ( \psi _{L} , \psi _{R} ) ^{T}$, the modified Dirac equation \eqref{Dirac_eq_spacelike_axial} splits into two Weyl-type equations with opposite axial couplings. For a general constant axial background
\begin{align}
    b _{\mu} = ( b _{0}, \mathbf{b} ) \qquad \text{where} \qquad \mathbf{b} = (b _{x} , b _{y} , b _{z} ) , 
\end{align}
the equations of motion read
\begin{align}
    i \hbar c \, \sigma ^{\mu} \partial _{\mu} \, \psi _{R} \; - \; \left( b _{0} - \mathbf b\!\cdot\!\boldsymbol{\sigma} \right) \psi _{R} &=0 , \label{Weyl_R_general} \\ i \hbar  c \, \bar{\sigma} ^{\mu} \partial _{\mu} \, \psi _{L} \; + \; \left( b _{0} + \mathbf{b} \cdot \boldsymbol{\sigma} \right) \psi _{L} &=0 .    \label{Weyl_L_general}
\end{align}
These equations make explicit that the axial background couples with opposite sign to the two chiral components of the Dirac field. The timelike component $b _{0}$ produces an energy-like shift that is identical for all spatial directions, while the spacelike components $\mathbf{b}$ act anisotropically through their coupling to the Pauli matrices. We note that while the Weyl representation for the gamma matrices is explicitly adopted here to simplify the algebraic manipulation of the $\gamma^5$ matrix, all physical observables—including the boundary quantization, the phase shift $\delta_\nu(k_z)$, and the final Casimir energy density—are strictly representation-independent. Any change of the spinor basis corresponds to a unitary transformation $U$ in the Clifford algebra, under which both the modified Dirac equation and the MIT bag boundary condition transform covariantly, preserving the eigenvalues of the boundary-value problem.

In momentum space, this structure admits a natural interpretation in terms of chirality-dependent shifts of the fermionic spectrum. For spacelike backgrounds, the axial coupling effectively modifies the momentum as $\mathbf{p} \to \mathbf{p} \mp \mathbf{b}$, with opposite signs for right- and left-handed modes, whereas a nonvanishing $b _{0}$ shifts the energy eigenvalues in opposite directions for the two chiralities. In unbounded space, both effects can often be absorbed into a redefinition of the energy and momentum variables and therefore do not lead to observable consequences.

In the presence of boundaries, however, this redefinition is obstructed by the quantization of the longitudinal modes. As a result, the impact of Lorentz violation becomes sensitive to the orientation of the axial background relative to the confining geometry. In particular, only those components of $\mathbf{b}$ that project along the direction of confinement can modify the discrete spectrum, while components parallel to the plates merely shift the continuous transverse momenta and drop out after the standard subtraction of the free-space vacuum energy. This mechanism underlies the subsequent analysis of the fermionic Casimir effect and explains why genuine Lorentz-violating corrections to the vacuum energy originate exclusively from background components aligned with the confinement direction.

%---------------- Spectrum mode --------------

\section{Mode decomposition and bulk spectrum} \label{mode_section}

In this section we construct the mode decomposition of the Dirac field in the presence of an axial Lorentz-violating background and determine the corresponding bulk dispersion relations. Owing to the different physical roles played by the temporal and spatial components of the background vector, we analyze separately the cases of timelike and spacelike axial configurations.

\subsection{Geometry, boundary conditions, and spinor structure} \label{subsec:geometry_bc_fermion}

We consider a planar waveguide geometry defined by two parallel plates located at $z=0$ and $z=L$, while the system remains unbounded and translationally invariant along the transverse directions $(x,y)$. The Dirac field is therefore confined to the region
\begin{align}
    0 < z < L \qquad \text{and} \qquad (x,y) \in \mathbb{R} ^{2} , \label{region_fermion_waveguide}
\end{align}
which preserves continuous momentum labels parallel to the plates and leads to a discrete spectrum only in the longitudinal direction.

Confinement at the plates is implemented through MIT bag boundary conditions \cite{MiltonBag,BordagBag,ElizaldeKirsten}, which ensure the vanishing of the normal component of the fermionic current at the boundaries. These conditions take the covariant form
\begin{align}
    \left( 1 + i \gamma ^{\mu} n _{\mu} \right) \psi \big| _{x \in \Sigma} = 0  ,    \label{MIT_general}
\end{align}
where $n _{\mu}$ is the outward normal to the boundary $\Sigma$. For the present geometry one has $n _{\mu} = (0,0,0,-1)$ at $z=0$ and $n _{\mu} = (0,0,0,+1)$ at $z=L$, leading to
\begin{align}
    ( 1 - i \gamma ^{3} ) \psi \big| _{z=0} = 0 , \qquad ( 1 + i \gamma ^{3} ) \psi \big| _{z=L} = 0 . \label{MIT_plates}
\end{align}
In the chiral (Weyl) representation, the matrix $\gamma ^{3}$ is off-diagonal, so that the boundary conditions intrinsically couple left- and right-handed components of the Dirac spinor. Writing the longitudinal spinor as
\begin{align}
    \Phi (z) = \begin{pmatrix} \phi _{L} (z) \\ \phi _{R} (z) \end{pmatrix} , \label{PhiLR_b0}
\end{align}
the MIT conditions imply the relations
\begin{align}
    \phi _{L} (0) = i\sigma ^{3} \phi _{R} (0) , \qquad \phi _{L} (L) = - i \sigma ^{3} \phi _{R} (L) , \label{MIT_LR_relations}
\end{align}
which will be used in the following to determine the allowed longitudinal modes.

These boundary conditions are independent of the specific orientation of the axial background and therefore provide a common starting point for the analysis of both timelike and spacelike Lorentz-violating configurations. The distinction between these cases enters exclusively through the bulk equations of motion and the resulting dispersion relations, which are discussed next. The assumption of a strictly constant and homogeneous axial vector ($\partial_\alpha b_\mu = 0$) is standard within the Standard-Model Extension framework, where such backgrounds correspond to cosmic vacuum expectation values smoothed out by inflation on laboratory scales. Phenomenologically, if an explicit spatial dependence perpendicular to the plates were introduced, such as $b_\mu(z)$, the translational invariance along the confinement axis would be broken. Consequently, the longitudinal momentum $k_z$ would no longer be a conserved quantum number, and the field equations would map onto a Schrödinger-like equation with a $z$-dependent potential. While this would prevent a simple plane-wave ansatz and lead to a non-uniform vacuum energy density $\mathcal{E}_{\mathrm{Cas}}(z)$ between the boundaries, the constant background chosen here isolates the pure geometric selection rules of the topological vector field.

Furthermore, it is important to emphasize that the exact cancellation of the transverse components $b_x$ and $b_y$ from the final Casimir energy density is a geometric consequence of the infinite-plate approximation. Since the transverse dimensions are unbounded, the continuous momenta $(k_x, k_y) \in \mathbb{R}^2$ can be freely shifted during the integration process. In realistic setups involving finite plates, lateral boundaries, or compactified transverse topologies, the continuous translation invariance is lost, and the parallel components of the axial vector would explicitly contribute to the physical vacuum forces due to edge and finite-size quantization effects.

\subsection{Timelike axial background} \label{subsec:timelike_modes}

We begin with the case of a purely timelike axial background,
\begin{align}
    b _{\mu} = (b _{0},0,0,0),
\end{align}
for which the axial Dirac equation reduces to two decoupled Weyl-type equations with opposite energy shifts, as discussed in Sec.~II. Owing to translational invariance parallel to the plates, the fermionic field can be expanded as a plane wave in $(t,x,y)$,
\begin{align}
    \psi (\mathbf{r},t) = e ^{-iEt/\hbar} \, e ^{i(k _{x} x + k _{y} y)} \, \Phi (z) , \label{mode_ansatz_b0}
\end{align}
where $\Phi (z)$ is a four-component spinor encoding the longitudinal dependence and is decomposed as in Eq.~\eqref{PhiLR_b0}.

Substitution of the ansatz in Eq. \eqref{mode_ansatz_b0} into the chiral Eqs. \eqref{Weyl_R_general}-\eqref{Weyl_L_general}, and introducing $\mathbf{k}_\perp=(k_x,k_y)$, leads to the first-order differential equations
\begin{align}
    \Big[ (E + b _{0} ) I - \hbar c ( \sigma ^{1} k _{x} + \sigma ^{2} k _{y} ) \Big] \phi _{R} (z) + i \hbar c \, \sigma ^{3} \frac{d}{dz} \phi _{R} (z) &=0 , \label{R_z_eq} \\ \Big[ ( E - b _{0} ) I + \hbar c ( \sigma ^{1} k _{x} + \sigma ^{2} k _{y} ) \Big] \phi _{L} (z) - i \hbar c \, \sigma ^{3} \frac{d}{dz} \phi _{L} (z) &=0 .  \label{L_z_eq}
\end{align}
Within the slab $ 0 < z < L$, both chiral components are superpositions of forward- and backward-propagating longitudinal modes,
\begin{align}
    \phi _{R} (z) = r _{+} e ^{ik _{z} z} + r _{-} e ^{-ik _{z} z} , \qquad \phi _{L} (z) = \ell _{+} e ^{ik _{z} z} + \ell _{-} e ^{- i k _{z} z} ,
\end{align}
with constant two-spinors $r _{\pm}$ and $\ell _{\pm}$. The MIT bag boundary conditions at $z=0$ and $z=L$, given by Eq. \eqref{MIT_LR_relations}, relate left- and right-handed components at the boundaries. Evaluating the mode expansions at $z=0$ and $z=L$ yields
\begin{align}
    \ell _{+} + \ell _{-} &= i \sigma ^{3} (r_{+} + r_{-} ) , \label{BC0_detail} \\ \ell _{+} e ^{ik _{z} L} + \ell _{-} e ^{-i k _{z} L }  &= - i \sigma ^{3} \left( r_{+} e ^{i k _{z} L } + r _{-} e ^{- i k _{z} L } \right) .  \label{BCL_detail}
\end{align}
The amplitudes are further constrained by the bulk equations of motion. Inserting the plane-wave solutions into Eqs.~\eqref{R_z_eq} and \eqref{L_z_eq} shows that each longitudinal exponential $e ^{\pm i k _{z} z}$ is an eigenmode provided
\begin{align}
    \Big[ ( E + b _{0} ) I - \hbar c ( \sigma ^{1} k _{x} + \sigma ^{2} k _{y} ) \mp \hbar c \, k _{z} \sigma ^{3} \Big] r _{\pm} &=0 , \\ \Big[ ( E - b _{0} ) I + \hbar c ( \sigma ^{1} k _{x} + \sigma ^{2} k _{y} ) \pm \hbar c \, k _{z} \sigma ^{3} \Big] \ell _{\pm} &=0.
\end{align}
Using these relations to eliminate $\ell _{\pm}$ from Eqs.~\eqref{BC0_detail}-\eqref{BCL_detail} yields a homogeneous linear system for the independent amplitudes $r _{\pm}$. The condition for nontrivial solutions reduces to the scalar phase relation
\begin{align}
    e ^{2 i k _{z} L} = \frac{ b _{0} - i \hbar c \, k _{z} }{ b _{0} + i \hbar c \, k _{z} } ,
\end{align}
which has unit modulus and can therefore be written as
$e ^{2 i k _{z}L} = e ^{- 2 i \delta _{b_{0}} (k _{z} ) }$, with phase shift
\begin{align}
    \delta _{b_{0}} (k _{z} ) = \arctan \left( \frac{\hbar c \, k _{z} }{b _{0}} \right) .
\end{align}
The allowed longitudinal momenta are then determined by
\begin{align}
    2 k _{z} L + 2 \delta _{b_{0}} ( k _{z} ) = 2n \pi , \qquad n=0,1,2, \ldots ,
\end{align}
or, equivalently,
\begin{align}
    k _{z} L + \arctan \left( \frac{\hbar c \, k _{z} }{b _{0} } \right) = n \pi . \label{kz_b0} 
\end{align}
In the Lorentz-symmetric limit $b _{0} \to 0$, the phase shift approaches $\pi/2$ and the standard half-integer spectrum of the massless MIT waveguide is recovered.

Finally, inserting the discrete solutions $k _{z} ^{(n)}$ into the bulk dispersion relation yields the confined mode energies,
\begin{align}
    E ^{(\pm,s)} _{n,\mathbf{k} _{\perp} } = \pm \hbar  c \, \sqrt{k _{\perp} ^{2} + \big[ k _{z} ^{(n)} \big] ^{2} } + s \, b _{0} , \qquad s = \pm 1 ,
\end{align}
which display the expected chirality-dependent energy shift induced by the timelike axial background.

\subsection{Spacelike axial background}
\label{subsec:spacelike_modes}

We now turn to the case of a purely spacelike axial background,
\begin{align}
    b_{\mu} = (0,\mathbf{b}) \qquad \text{where} \qquad \mathbf{b} = (b_{x},b_{y},b_{z}) ,
\end{align}
for which the axial coupling enters the Weyl equations as a momentum-like term with opposite sign for the two chiralities. As in the timelike case, translational invariance parallel to the plates allows a plane-wave decomposition in $(t,x,y)$,
\begin{align}
    \psi (\mathbf{r},t) = e ^{-iEt/\hbar} \, e ^{i(k _{x} x + k _{y} y)} \, \Phi (z) ,
    \label{mode_ansatz_spacelike}
\end{align}
where the longitudinal dependence is encoded in the four-component spinor
\begin{align}
    \Phi (z) =
    \begin{pmatrix}
        \phi _{L} (z) \\
        \phi _{R} (z)
    \end{pmatrix} .
    \label{PhiLR_spacelike}
\end{align}
Substitution of the ansatz in Eq. \eqref{mode_ansatz_spacelike} into the Weyl Eqs. \eqref{Weyl_R_general}-\eqref{Weyl_L_general}, and introducing $\mathbf{k} _{\parallel} = (k _{x},k _{y})$ and $\mathbf{b} _{\parallel} = (b _{x},b _{y})$, yields the first-order equations along the longitudinal direction:
\begin{align}
    \Big[
        E I
        - \hbar c ( \sigma ^{1} k _{x} + \sigma ^{2} k _{y} )
        - ( \sigma ^{1} b _{x} + \sigma ^{2} b _{y} )
        - b _{z} \sigma ^{3}
    \Big] \phi _{R} (z)
    + i \hbar c \, \sigma ^{3} \frac{d}{dz} \phi _{R} (z)
    &= 0 ,
    \label{R_eq_spacelike_z}
    \\
    \Big[
        E I
        + \hbar c ( \sigma ^{1} k _{x} + \sigma ^{2} k _{y} )
        + ( \sigma ^{1} b _{x} + \sigma ^{2} b _{y} )
        + b _{z} \sigma ^{3}
    \Big] \phi _{L} (z)
    - i \hbar c \, \sigma ^{3} \frac{d}{dz} \phi _{L} (z)
    &= 0 .
    \label{L_eq_spacelike_z}
\end{align}
Within the slab $0 < z < L$, both chiral components are written as superpositions of forward- and backward-propagating longitudinal waves,
\begin{align}
    \phi _{R} (z)
    =
    r _{+} e ^{i k _{z} z}
    +
    r _{-} e ^{- i k _{z} z}
    \qquad \text{and} \qquad
    \phi _{L} (z)
    =
    \ell _{+} e ^{i k _{z} z}
    +
    \ell _{-} e ^{- i k _{z} z} ,
\end{align}
with constant two-spinors $r _{\pm}$ and $\ell _{\pm}$.

The amplitudes are constrained by the bulk equations of motion. Inserting the plane-wave solutions into Eqs.~\eqref{R_eq_spacelike_z} and \eqref{L_eq_spacelike_z} shows that each exponential $e ^{\pm i k _{z} z}$ is an eigenmode provided
\begin{align}
    \Big[
        E I
        - \hbar c ( \sigma ^{1} k _{x} + \sigma ^{2} k _{y} )
        - ( \sigma ^{1} b _{x} + \sigma ^{2} b _{y} )
        - b _{z} \sigma ^{3}
        \mp \hbar c \, k _{z} \sigma ^{3}
    \Big] r _{\pm} &= 0 ,
    \label{r_pm_spacelike}
    \\
    \Big[
        E I
        + \hbar c ( \sigma ^{1} k _{x} + \sigma ^{2} k _{y} )
        + ( \sigma ^{1} b _{x} + \sigma ^{2} b _{y} )
        + b _{z} \sigma ^{3}
        \pm \hbar c \, k _{z} \sigma ^{3}
    \Big] \ell _{\pm} &= 0 .
    \label{l_pm_spacelike}
\end{align}
Imposing now the MIT bag boundary conditions one obtains:
\begin{align}
    \ell _{+} + \ell _{-}
    &=
    i \sigma ^{3} ( r _{+} + r _{-} ) ,
    \label{BC0_spacelike}
    \\
    \ell _{+} e ^{i k _{z} L} + \ell _{-} e ^{- i k _{z} L}
    &=
    - i \sigma ^{3}
    \left(
        r _{+} e ^{i k _{z} L}
        +
        r _{-} e ^{- i k _{z} L}
    \right) .
    \label{BCL_spacelike}
\end{align}
Using the bulk relations \eqref{l_pm_spacelike} to eliminate $\ell _{\pm}$ from Eqs.~\eqref{BC0_spacelike}-\eqref{BCL_spacelike}, one arrives at a homogeneous linear system for the independent amplitudes $r _{\pm}$. The condition for nontrivial solutions reduces to the scalar phase relation
\begin{align}
    e ^{2 i k _{z} L}
    =
    \frac{ b _{z} - i \hbar c \, k _{z} }
         { b _{z} + i \hbar c \, k _{z} } ,
    \label{phase_bz}
\end{align}
which has unit modulus and can therefore be written as $e ^{2 i k _{z} L} = e ^{- 2 i \delta _{b_{z}} ( k _{z} ) }$, with phase shift
\begin{align}
    \delta _{b_{z}} ( k _{z} )
    =
    \arctan \left( \frac{ \hbar c \, k _{z} }{ b _{z} } \right) .
\end{align}
The allowed longitudinal momenta are then determined by
\begin{align}
    2 k _{z} L + 2 \delta _{b_{z}} ( k _{z} )
    = 2 n \pi , \qquad n = 0,1,2,\ldots ,
\end{align}
or, equivalently,
\begin{align}
    k _{z} L
    +
    \arctan \left( \frac{ \hbar c \, k _{z} }{ b _{z} } \right)
    =
    n \pi . \label{kz_bz}
\end{align}
In the Lorentz-symmetric limit $b _{z} \to 0$, the standard half-integer MIT spectrum is recovered.

Finally, inserting the discrete solutions $k _{z} ^{(n)}$ into the bulk dispersion relation yields the confined mode energies,
\begin{align}
    E ^{(\pm,s)} _{n,\mathbf{k} _{\parallel}}
    =
    \pm \hbar c
    \sqrt{
        \left|
            \mathbf{k} _{\parallel}
            +
            s \, \frac{ \mathbf{b} _{\parallel} }{ \hbar c }
        \right| ^{2}
        +
        \big[ k _{z} ^{(n)} \big] ^{2}
    } ,
    \qquad
    s = \pm 1 ,
\end{align}
which makes explicit that the parallel components $\mathbf{b} _{\parallel}$ enter only through a continuous shift of the transverse momenta, while the normal component $b _{z}$ induces a genuine modification of the longitudinal quantization via the phase shift $\delta _{b_{z}}$.

%----------------- Casimir energy ---------

\section{Vacuum energy and Casimir interaction}
\label{sec:casimir_unified}

The fermionic Casimir effect in the present geometry is governed by the vacuum energy associated with the normal modes of the axial Dirac field confined between two parallel plates. Once the allowed longitudinal momenta have been determined from the quantization conditions \eqref{kz_b0} and \eqref{kz_bz}, the vacuum energy can be written formally as a sum over the normal modes. For the timelike case, the two branches labeled by $s=\pm1$ contribute through the energies
\begin{align}
    E ^{(\pm,s)} _{n,\mathbf{k} _{\perp} } = \pm \hbar  c \, \sqrt{k _{\perp} ^{2} + \big[ k _{z} ^{(n)} \big] ^{2} } + s \, b _{0} ,
\end{align}
whereas for the spacelike case the branches differ by a shift of the transverse momentum,
\begin{align}
    E ^{(\pm,s)} _{n,\mathbf{k} _{\parallel}}
    =
    \pm \hbar c
    \sqrt{
        \left|
            \mathbf{k} _{\parallel}
            +
            s \, \frac{ \mathbf{b} _{\parallel} }{ \hbar c }
        \right| ^{2}
        +
        \big[ k _{z} ^{(n)} \big] ^{2}
    } .
\end{align}
In both situations, after summing over the two branches and taking into account the shift invariance of the transverse momentum integrals, the formal vacuum energy per unit area takes the unified form
\begin{align}
    \mathcal{E} _{0}
    =
    - 2
    \sum _{n}
    \int
    \frac{d ^{2}\mathbf{k} _{\perp}}{(2 \pi ) ^{2}}
    \,
    \hbar c
    \sqrt{
        k _{\perp} ^{2}
        +
        \big[ k _{z} ^{(n)} \big] ^{2}
    } ,
    \label{E0_per_area}
\end{align}
which is ultraviolet divergent and must therefore be regularized by subtracting the vacuum contribution of the unbounded geometry.

A crucial simplification of the present problem is that, although the microscopic origin of the spectral shift is different in the timelike and spacelike configurations, the resulting longitudinal quantization conditions have exactly the same functional structure. Indeed, for the timelike axial background one finds from Eq.~\eqref{kz_b0}
\begin{align}
    k _{z} L + \arctan \left(\frac{\hbar c\,k _{z} }{b _{0}} \right) = n \pi ,
\end{align}
whereas for the spacelike axial background normal to the plates one obtains from Eq.~\eqref{kz_bz}
\begin{align}
    k _{z} L + \arctan \left( \frac{\hbar c \, k _{z} }{b _{z} } \right) = n \pi .
\end{align}
Therefore, both cases can be treated simultaneously by introducing the effective Lorentz-violating parameter
\begin{align}
    b _{\ast} = \begin{cases}
        b _{0} , & \text{timelike axial background},\\[3pt]
        b _{z} , & \text{spacelike axial background normal to the plates},
    \end{cases}
\end{align}
together with the inverse-length scale
\begin{align}
    \nu \equiv \frac{|b _{\ast}|}{\hbar c}.
    \label{nu_unified_def}
\end{align}
In terms of this parameter, the longitudinal spectrum is determined by the unified quantization condition
\begin{align}
    k _{z} L + \arctan \left( \frac{k _{z}}{\nu} \right) = n \pi ,  \qquad n = 0 , 1 , 2 , \ldots .    \label{unified_quantization_condition}
\end{align}
This representation makes explicit that the Casimir problem depends only on the dimensionless combination $\nu L=|b _{\ast}|L/(\hbar c)$.

It is important to emphasize that the exact cancellation of the transverse components $b_x$ and $b_y$ from the final Casimir energy density is a geometric consequence of the infinite-plate approximation. Since the transverse dimensions are unbounded, the continuous momenta $(k_x, k_y) \in \mathbb{R}^2$ can be freely shifted during the integration process. In realistic setups involving finite plates, lateral boundaries, or compactified transverse topologies, the continuous translation invariance is lost, and the parallel components of the axial vector would explicitly contribute to the physical vacuum forces due to edge and finite-size quantization effects.

We note that, within this infinite geometric framework, the signs of the individual components $b_0$ and $b_z$ do not influence the quantization conditions. A change in sign, such as $b_z \to -b_z$ (or $b_0 \to -b_0$), merely interchanges the discrete spectra of the two independent chiral sectors. Since the total vacuum energy accounts for the sum over the complete Dirac multiplet, the final expression depends strictly on the magnitude $|b_{\ast}|$ via the scale $\nu$.

Before proceeding with the regularization, it is useful to make explicit why the components of the spacelike background parallel to the plates do not affect the Casimir interaction. From the bulk dispersion relation in the spacelike case,
\begin{align}
    E ^{(\pm,s)} _{n,\mathbf{k} _{\parallel} } = \pm \hbar c  \sqrt{  \left| \mathbf{k} _{\parallel} + s \, \frac{\mathbf{b} _{\parallel} }{\hbar c} \right| ^{2}  + \big[ k _{z} ^{(n)} \big] ^{2}  } ,
\end{align}
one sees that $\mathbf{b} _{\parallel} = (b _{x},b _{y})$ enters only through a constant shift of the continuous transverse momentum. Accordingly, the vacuum-energy density contains integrals of the form
\begin{align}
    \int \frac{d ^{2} \mathbf{k} _{\parallel} }{(2 \pi ) ^{2} } \, f \left(  \left| \mathbf{k} _{\parallel} + s \, \frac{\mathbf{b} _{\parallel}}{\hbar c} \right| ^{2}  \right) ,
\end{align}
for each branch $s = \pm$. Since the integration domain is the entire plane $\mathbb{R} ^{2}$, we can see that by a simple redefinition of the variable $\mathbf{k} _{\parallel}$ the integral is in fact independent of $\mathbf{b} _{\parallel}$. Here we would like to highlight an important difference between  the results obtained for the Casimir energy in this model, with the another one that presents CPT-even Lorentz violation analyzed in \cite{PhysRevD.99.085012}. While in this CPT-odd the $b_	\parallel$ dropping out entirely, in CPT-even it produces a multiplicative factor in the Casimir energy.

%---------------------------

\subsection{Density of states and phase-shift representation}

The unified quantization condition in Eq. \eqref{unified_quantization_condition} can be written in terms of the phase shift:
\begin{align}
    \delta _{\nu} (k _{z}) = \arctan \left( \frac{k _{z}}{\nu} \right) ,  \label{delta_unified}
\end{align}
so that the counting function becomes
\begin{align}
    n(k _{z}) = \frac{1}{\pi} \Big[ k _{z} L + \delta _{\nu} ( k _{z} ) \Big]  .
\end{align}
Differentiation with respect to $k _{z}$ gives the exact density of longitudinal states,
\begin{align}
    \rho ( k _{z}) \equiv \frac{dn}{dk _{z}} = \frac{L}{\pi} + \frac{1}{\pi} \frac{d \delta _{\nu} (k _{z})}{dk _{z}} ,    \label{rho_unified}
\end{align}
with
\begin{align}
    \frac{d \delta _{\nu}}{dk _{z}} = \frac{\nu}{k _{z} ^{2} + \nu ^{2} } .   \label{delta_prime_unified}
\end{align}

Accordingly, the sum over discrete longitudinal modes may be replaced by an integral weighted with the density of states,
\begin{align}
    \sum _{n} f \big( k _{z} ^{(n)} \big)
    \;\longrightarrow\;
    \int _{0} ^{\infty} d k _{z} \, \rho ( k _{z} ) \, f ( k _{z} ) .
    \label{sum_to_dos}
\end{align}
The first term in Eq. \eqref{rho_unified}, proportional to $L/\pi$, reproduces the density of states of the unbounded system and is removed by the standard Casimir subtraction. The remaining contribution is entirely determined by the phase shift induced by the plates in the presence of the effective parameter $b _{\ast}$. Accordingly, the Casimir energy density can be written as:
\begin{align}
    \mathcal{E} _{\mathrm{Cas}}
    =
    - \frac{2 \hbar c}{\pi}
    \int
    \frac{d ^{2}\mathbf{k} _{\perp}}{( 2 \pi ) ^{2}}
    \int _{0} ^{\infty}
    d k _{z}
    \,
    \frac{d \delta _{\nu} ( k _{z} ) }{d k _{z} }
    \,
    \sqrt{k _{\perp} ^{2} + k _{z} ^{2}} .
    \label{Ecas_phase_start}
\end{align}
The overall minus sign in Eq.~\eqref{Ecas_phase_start} is characteristic of the fermionic vacuum energy representation, reflecting the anticommuting nature of the field. Throughout this work, a negative renormalized Casimir energy density $\mathcal{E}_{\mathrm{Cas}} < 0$ implies an attractive mechanical pressure between the plates, defined by:
\begin{equation}
    \mathcal{P}_C = -\frac{\partial (\mathcal{E}_{\mathrm{Cas}} L)}{\partial L},
\end{equation}
which is consistent with the standard massless fermionic Casimir effect. Introducing
\begin{align}
    \omega (k _{\perp}, k _{z}) \equiv \sqrt{ k _{\perp} ^{2} + k _{z} ^{2}} ,
\end{align}
one may integrate by parts with respect to $k _{z}$ to obtain the equivalent phase-shift representation
\begin{align}
    \mathcal{E} _{\mathrm{Cas}} = \frac{2 \hbar c}{\pi} \int \frac{ d ^{2} \mathbf{k} _{\perp}}{( 2 \pi ) ^{2}} \int _{0} ^{\infty} d k _{z} \, \delta _{\nu} ( k _{z} ) \, \frac{k _{z}}{\sqrt{ k _{\perp} ^{2} + k _{z} ^{2}}} ,
    \label{Ecas_after_ibp_unified}
\end{align}
where the boundary term, being independent of the plate separation, is absorbed into the same renormalization prescription.

%----------------------------------

\subsection{Derivation of the closed logarithmic representation}

In order to convert the phase-shift representation into a closed logarithmic one, it is useful to introduce the spectral function associated with the longitudinal quantization condition. Since
\begin{align}
    k _{z} L + \delta _{\nu} ( k _{z} ) = n \pi ,
\end{align}
one may equivalently write
\begin{align}
    e ^{2i[k _{z} L + \delta _{\nu} (k _{z})]} =  1 .
\end{align}
Using
\begin{align}
    e ^{2i \delta _{\nu} ( k _{z} ) } = \frac{ \nu + i k _{z}}{ \nu - i k _{z}} ,
\end{align}
one is naturally led to define
\begin{align}
    F ( k _{z} ) = 1 + e ^{2ik _{z}L} \, \frac{ \nu + i k _{z}}{ \nu - i k _{z}} ,    \label{spectral_function_F_unified}
\end{align}
whose zeros determine the allowed longitudinal modes. The renormalized vacuum energy can then be written in the standard argument-principle form
\begin{align}
    \mathcal{E} _{\mathrm{Cas}} = - 2 \hbar c \int \frac{d ^{2} \mathbf{k} _{\perp}}{ ( 2 \pi ) ^{2} } \frac{1}{2 \pi i} \int _{\mathcal{C} } d k _{z} \, \omega ( k _{\perp},k _{z} ) \, \frac{d}{d k _{z}} \ln F ( k _{z} ) ,    \label{argument_principle_energy_unified}
\end{align}
where $\mathcal{C}$ is a contour enclosing the positive real zeros of $F ( k _{z} )$.

After subtracting the free-space contribution and deforming the contour to the imaginary axis, the branch-cut contribution can be evaluated in the standard way. For MIT bag boundary conditions, one arrives at the compact representation
\begin{align}
    \mathcal{E} _{\mathrm{Cas}} ( b _{\ast} ) = - \frac{\hbar c}{\pi ^{2}} \int _{0} ^{\infty} dk \; k ^{2} \, \ln \left( 1 + e ^{-2L \sqrt{k ^{2} + \nu ^{2}}} \right) , \qquad \nu = \frac{|b _{\ast}|}{\hbar c} , \label{Ecas_unified_master}
\end{align}
which is the desired closed logarithmic form. The Eq. \eqref{Ecas_unified_master} applies equally to the timelike case ($ b _{\ast} = b _{0}$) and to the spacelike case with background component normal to the plates ($b _{\ast} = b _{z}$). It shows that Lorentz violation enters the fermionic Casimir interaction in exactly the same way as an effective mass scale. It is worth mentioning that this mathematical equivalence to a massive fermionic system can be traced back to the transcendental mode quantization. In a standard massive Dirac field with mass $m$ confined by MIT bag boundaries without Lorentz violation, the boundary-induced phase shift is governed by $\arctan(k_z/m)$. In our framework, the unified parameter $\nu = |b_{\ast}|/(\hbar c)$ plays an identical role in the transcendental equation, yielding the mass-like structure in Eq.~\eqref{Ecas_unified_master}, although the bulk dispersion relation remains fundamentally distinct due to the Lorentz-violating chiral splitting.

In particular, the Lorentz-symmetric limit is recovered smoothly by taking $\nu \to 0$, yielding: \cite{Milton2001,Bordag2009}
\begin{align}
    \mathcal{E} _{\mathrm{Cas}} ( \nu \to 0 ) = - \frac{7 \pi ^{2}}{2880} \, \frac{\hbar c}{L ^{3}} , \label{Ecas_massless_limit}
\end{align}
namely the standard massless Dirac result for MIT bag boundary conditions.

\begin{figure}[h!]
    \centering
    \includegraphics[width=0.6\textwidth]{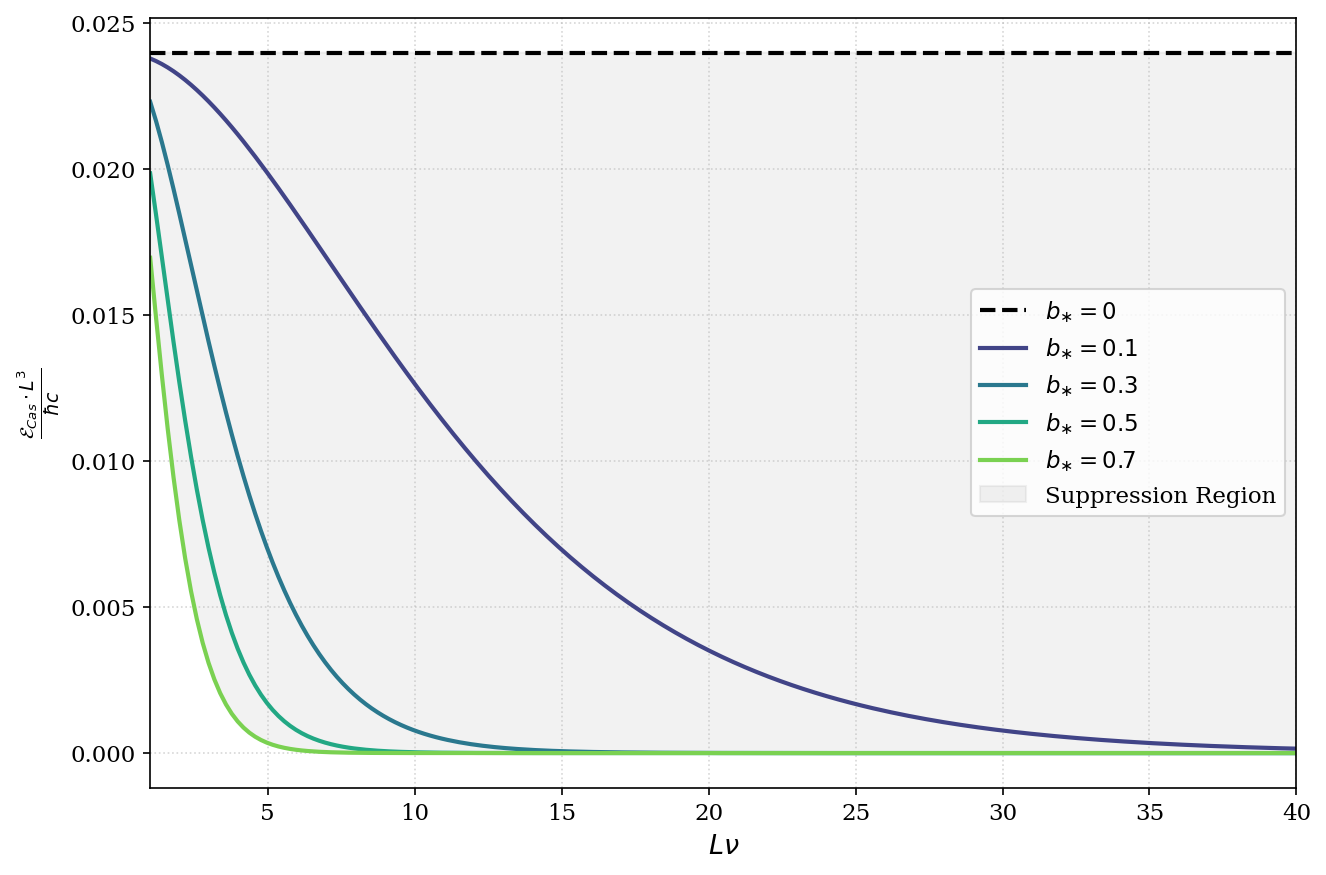}
    \caption{Dimensionless Casimir energy density as a function of the plate separation $L \nu$ for different Lorentz-violating parameters $b_{\ast}$. The dashed black line represents the standard Lorentz-invariant Dirac fermion limit ($b_{\ast} = 0$). The shaded gray area indicates the suppression region, where the broken symmetry leads to a significant reduction in vacuum energy compared to the power-law decay.}
    \label{fig:casimir_energy_plot}
\end{figure}

The numerical evaluation of Eq. \eqref{Ecas_unified_master} reveals the critical role of the Lorentz-violating parameter $b_{\ast}$ in modulating the vacuum energy density. In Fig. \ref{fig:casimir_energy_plot}, we present the energy scaling function $\mathcal{E}_{\mathrm{Cas}} \cdot L^3 / (\hbar c)$ as a function of the plate separation $L \nu$. For the Lorentz-invariant case ($b_{\ast} = 0$), the scaling function remains constant, recovering the expected power-law decay $\mathcal{E} \propto L^{-3}$ characteristic of massless Dirac fermions. However, as $b_{\ast}$ increases, a clear departure from this regime is observed. The interaction enters a suppression regime, where the vacuum fluctuations are exponentially damped as the dimensionless product $L\nu$ grows. This behavior suggests that the axial vector $b_{\ast}$ acts as an effective mass gap, filtering out long-wavelength modes that would otherwise contribute to the Casimir effect.

\begin{figure}[h!]
    \centering
    \includegraphics[width=0.6\textwidth]{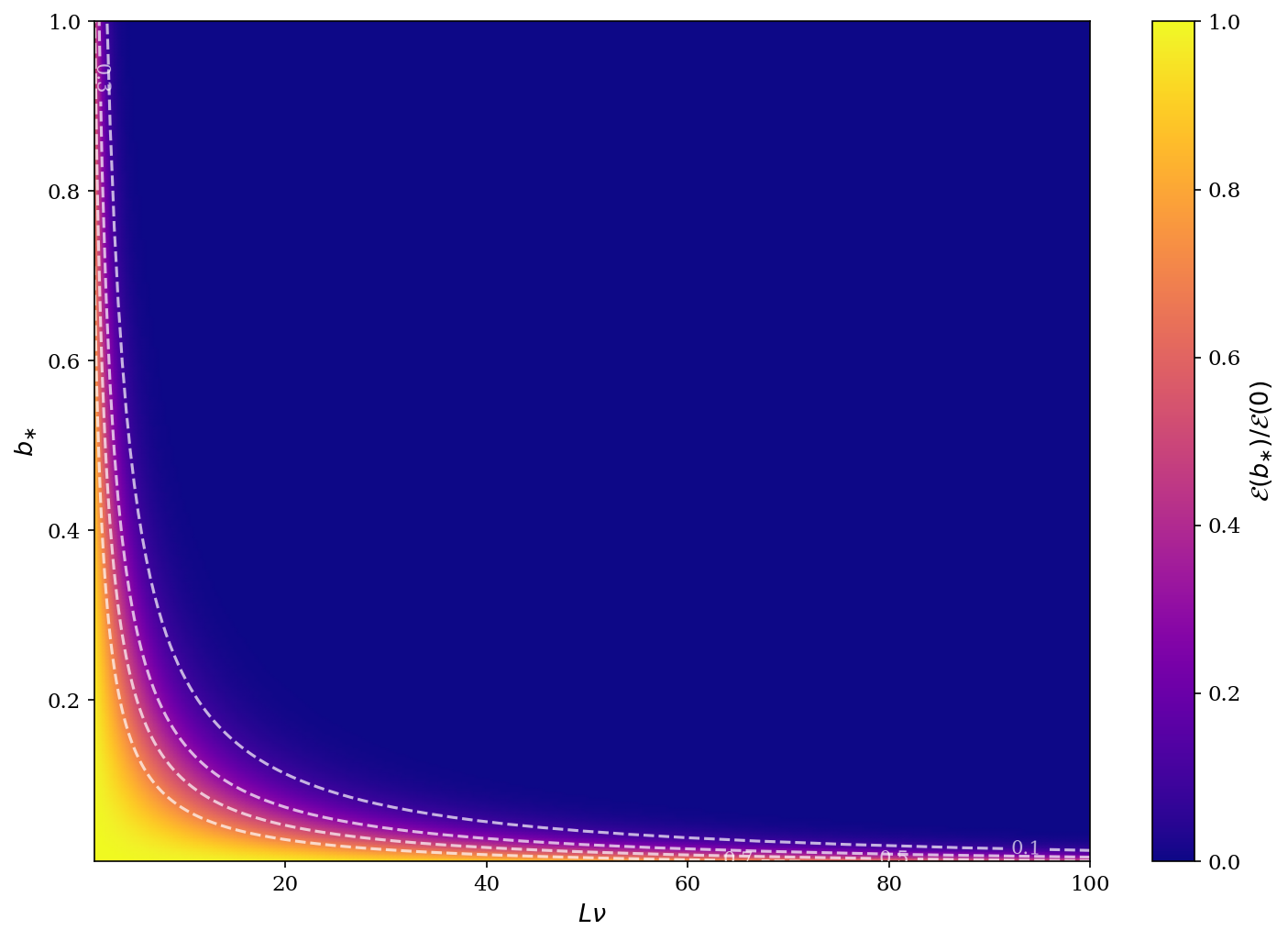}
    \caption{Heatmap of the normalized Casimir energy density $\mathcal{E}(b_{\ast}) / \mathcal{E}(0)$ as a function of the plate separation $L\nu$ and the Lorentz-violating parameter $b_{\ast}$. The color gradient represents the suppression ratio, where 1.0 (yellow) corresponds to the standard Lorentz-invariant Dirac limit and 0 (dark purple) indicates total vacuum energy suppression. The dashed white isolines highlight the thresholds where the energy is reduced to 90\%, 50\%, and 10\% of its original value.}
    \label{fig:casimir_energy_map}
\end{figure}

To further explore the interplay between the geometric and symmetry-breaking scales in Eq. \eqref{Ecas_unified_master}, we plot in Fig. \ref{fig:casimir_energy_map} a heatmap of the energy ratio $\mathcal{E}(b_{\ast}) / \mathcal{E}(0)$. This visualization provides a comprehensive overview of the parameter space, mapping the transition from the Dirac-like regime (where the ratio approaches unity) to the suppression zone (where the vacuum energy is significantly depleted). The contour lines follow hyperbolic trajectories defined by $L b_{\ast} \approx \text{const.}$, confirming that the suppression of the Casimir interaction is a universal function of the scaling parameter $\nu L$. These findings are particularly relevant for experimental setups involving Weyl semimetals, where the nodal separation can be tuned to modulate quantum vacuum effects.

%---------------- Weak Regime ------------------
\subsection{Weak Lorentz-violation regime: $\nu L \ll 1$}

To obtain the small-$\nu$ behavior, define
\begin{align}
    I(\nu)
    \equiv
    \int_0^\infty dk\;k^2\,
    \ln \left(
        1+e^{-2L\sqrt{k^2+\nu^2}}
    \right),
    \qquad
    \mathcal E_{\mathrm{Cas}}(b_\ast)
    =
    -\frac{\hbar c}{\pi^2}I(\nu).
    \label{I_unified_def}
\end{align}
Since the integrand depends on $\nu$ only through $\nu^2$, the expansion is naturally organized in powers of $\nu^2$. Differentiating with respect to $\nu^2$ gives
\begin{align}
    \frac{dI}{d(\nu^2)}
    =
    -L
    \int_0^\infty dk\;
    \frac{k^2}{\sqrt{k^2+\nu^2}}\,
    \frac{1}{e^{2L\sqrt{k^2+\nu^2}}+1}.
    \label{Iprime_unified}
\end{align}
Evaluating at $\nu=0$ yields
\begin{align}
    \left.\frac{dI}{d(\nu^2)}\right|_{\nu=0}
    &=
    -L\int_0^\infty dk\;\frac{k}{e^{2Lk}+1}
    =
    -\frac{\pi^2}{48\,L},
\end{align}
while
\begin{align}
    I(0)
    =
    \int_0^\infty dk\;k^2\ln(1+e^{-2Lk})
    =
    \frac{7\pi^4}{2880\,L^3}.
\end{align}
Therefore, for $\nu L \ll 1$ the Casimir energy density takes the form
\begin{align}
    \mathcal E_{\mathrm{Cas}}(b_\ast)
    &=
    -\frac{7\pi^2}{2880}\,\frac{\hbar c}{L^3}
    +
    \frac{b_\ast^2}{48\,\hbar c}\,\frac{1}{L}
    +
    \mathcal{O} \Big(
        \frac{b_\ast^4}{(\hbar c)^3}\ln\!\frac{|b_\ast|L}{\hbar c}
    \Big).
    \label{Ecas_small_unified}
\end{align}
Here, the logarithmic correction originates from the small-momentum ($k \to 0$) asymptotic expansion of the integrand in Eq.~\eqref{I_unified_def}. The sub-leading non-analytic terms of the form $x^2 \ln x$ appearing in the regularized expansion of the integrand generate the characteristic $\nu^4 \ln(\nu L)$ factor when integrated near the lower limit. This shows that the leading Lorentz-violating correction is quadratic in the effective background parameter and weakens the magnitude of the attractive Casimir energy.

%------------ Strong Regime ----------------
\subsection{Strong Lorentz-violation regime: $\nu L \gg 1$}

In the opposite limit, the exponential factor in \eqref{Ecas_unified_master} is strongly suppressed, and one may use $\ln(1+e^{-x})\simeq e^{-x}$ for $x\gg1$. Keeping the leading contribution, one obtains
\begin{align}
    \mathcal E_{\mathrm{Cas}}(b_\ast)
    \simeq
    -\frac{\hbar c}{\pi^2}
    \int_0^\infty dk\;k^2\,
    e^{-2L\sqrt{k^2+\nu^2}}.
\end{align}
For $\nu L\gg1$, the dominant region is $k\ll\nu$, so that
\begin{align}
    \sqrt{k^2+\nu^2}
    =
    \nu+\frac{k^2}{2\nu}+\cdots .
\end{align}
Substituting this approximation gives
\begin{align}
    \mathcal E_{\mathrm{Cas}}(b_\ast)
    &\simeq
    -\frac{\hbar c}{\pi^2}
    e^{-2\nu L}
    \int_0^\infty dk\;k^2 e^{-(L/\nu)k^2}.
\end{align}
Using the Gaussian integral
\begin{align}
    \int_0^\infty dk\;k^2 e^{-a k^2}
    =
    \frac{\sqrt{\pi}}{4}a^{-3/2},
    \qquad a>0,
\end{align}
one finds
\begin{align}
    \mathcal E_{\mathrm{Cas}}(b_\ast)
    \simeq
    -\frac{\hbar c}{4\pi^{3/2}}
    \left(\frac{\nu}{L}\right)^{3/2}
    e^{-2\nu L},
    \qquad \nu L\gg1.
\end{align}
Restoring the original parameter $b_\ast$, this becomes
\begin{align}
    \mathcal E_{\mathrm{Cas}}(b_\ast)
    \simeq
    -\frac{1}{4\pi^{3/2}}
    \frac{|b_\ast|^{3/2}}{\hbar^{1/2}c^{1/2}}
    \frac{1}{L^{3/2}}
    \exp \left(
        -\frac{2|b_\ast|}{\hbar c}\,L
    \right),
    \qquad
    \frac{|b_\ast|}{\hbar c}\,L\gg1.
    \label{Ecas_large_unified}
\end{align}
Hence, large Lorentz-violating backgrounds suppress the Casimir interaction exponentially, in complete analogy with the decoupling produced by a large fermion mass.

%-------- Conclusions ------------------
\section{Conclusions}
\label{conclusions}

In this work we have investigated the fermionic Casimir effect for a Dirac field confined between two parallel plates with MIT bag boundary conditions in the presence of a CPT-odd axial Lorentz-violating background constant vector $b_{\mu}$. Starting from the modified Dirac equation, we analyzed the spectral problem in planar geometry and derived the exact quantization conditions for the longitudinal momentum for both timelike and spacelike configurations of the background vector. Our analysis shows that the impact of Lorentz violation on the Casimir interaction is governed by the orientation of the axial vector relative to the confinement direction.

For a purely timelike background, the axial coupling produces opposite energy shifts for the two chiral sectors and leads to a modified phase shift in the longitudinal quantization condition. For a purely spacelike background, we found that only the component normal to the plates affects the discrete spectrum. In contrast, the components parallel to the plates merely shift the continuous transverse momenta and therefore disappear from the Casimir energy after the standard subtraction of the free-space contribution. This establishes a clear geometric selection rule: only the projection of the axial background vector along the confinement direction generates a genuine Lorentz-violating correction to the vacuum interaction.

A central result of this work is that the timelike component $b_{0}$ and the normal spacelike component $b_{z}$ can be treated within a unified framework. In both cases, the longitudinal spectrum is governed by the same functional quantization condition once one introduces the effective parameter $b_{\ast}$ and the associated inverse-length scale $\nu=|b_{\ast}|/(\hbar c)$. This unification makes it possible to formulate the vacuum energy through a single density-of-states representation and to derive a closed logarithmic expression for the renormalized Casimir energy density. The resulting formula shows that Lorentz violation enters the Casimir problem in a way formally analogous to the role played by a mass parameter in the conventional massive MIT problem.

The closed representation obtained here also allows for a transparent analysis of the asymptotic regimes. In the Lorentz-symmetric limit, the standard massless fermionic Casimir energy is recovered. In the weak-background regime, the leading correction is quadratic in $b_{\ast}$ and scales as $L^{-1}$, indicating that Lorentz violation weakens the magnitude of the attractive Casimir interaction. In the opposite strong-background regime, the energy becomes exponentially suppressed, showing that large axial background vector effectively decouple the confined vacuum fluctuations from the plate separation scale.

At this point,  we would  like once more to emphasize the contrast in the present results with those obtained in Ref.~\cite{PhysRevD.99.085012} for a CPT-even aether-like Lorentz-violating background. Although both scenarios share a similar planar geometry and MIT bag boundary conditions, they exhibit fundamentally distinct qualitative behaviors due to the different nature of the couplings. In the CPT-even case, the components of the background vector parallel to the plates ($b_\parallel$) cannot be discarded; instead, they induce a non-trivial multiplicative rescaling of the transverse coordinates, explicitly modifying the overall magnitude of the Casimir energy density. In striking contrast, for the CPT-odd axial background investigated here, the parallel components $b_x$ and $b_y$ drop out entirely from the renormalized vacuum energy due to a continuous momentum shift allowed by the infinite-plate geometry. Furthermore, we remark that while the final expression for the energy density maps mathematically onto the standard Elizalde–Kirsten massive-MIT bag integral \cite{ElizaldeKirsten}, the core physical novelty of our development lies not in the final integral structure itself, but rather in the non-trivial mapping mechanism. We demonstrate that a CPT-odd chiral coupling, through a unified spectral framework, induces an effective mass-like gap $\nu = |b_*|/(\hbar c)$ exclusively from the timelike and normal spacelike projections, a feature completely absent in conventional vector perturbations.

Beyond its high-energy interpretation within the SME, the present theory also has a natural connection with condensed-matter physics. Effective relativistic descriptions involving axial-vector couplings arise naturally in materials hosting Weyl and Dirac quasiparticles. In particular, in Weyl semimetals the separation of Weyl nodes in momentum space, and more generally the breaking of inversion or time-reversal symmetry that distinguishes the two chiral sectors, can be described at low energies by terms formally analogous to the axial-vector coupling considered here. Within this correspondence, the spacelike part of $b_{\mu}$ is associated with the momentum-space separation of the Weyl nodes, while a timelike component may be interpreted as an energy offset between nodes of opposite chirality.

The possible relevance of the present results to condensed-matter systems should nevertheless be interpreted with some care. Our analysis concerns a relativistic fermionic vacuum subject to ideal boundaries, whereas in a Weyl semimetal the low-energy fermions are emergent quasiparticles propagating in a crystalline medium with a microscopic lattice structure. Consequently, the Casimir energy obtained here should not be viewed as a direct observable prediction for a real Weyl material. Nevertheless, the effective field-theory description of Weyl semimetals often involves axial-vector parameters formally analogous to the background $b_{\mu}$ considered here. From this perspective, our analysis highlights a simple spectral property of axial couplings under confinement: only the component of the effective axial vector along the confinement direction modifies the discrete mode quantization, whereas the transverse components are absorbed into shifts of the conserved momenta. This mechanism may provide qualitative insight into orientation-dependent finite-size effects in effective models of confined Weyl materials.

Finally, we stress that the exponential suppression of the Casimir interaction found in the strong-background regime ($\nu L \gg 1$) provides a clear signature that could be verified in realistic experimental setups. By engineering a thin-film heterostructure of thickness $L$ and dynamically tuning the effective parameter $b_\ast$ through mechanical strain or external fields \cite{PhysRevResearch.4.023106, Armitage2018, Grushin2012}, one could modulate the material from a power-law Dirac-like regime to an exponentially suppressed zone. This tunable fluctuation-induced force opens up interesting technological possibilities for the control of micro- and nano-mechanical devices using topological quantum materials \cite{Capasso:2007zz, Woods:2016vwu, PhysRevResearch.3.023201}.

Furthermore, our results offer a clear criterion to distinguish Lorentz-violating effects from other beyond the Standard Model (BSM) signatures. While conventional dark forces or isotropic BSM fields yield corrections that depend solely on the plate separation $L$, the axial background $b_\mu$ induces a strict spatial anisotropy governed by a geometric selection rule. Consequently, experimental setups monitoring sidereal variations of the Casimir force as the apparatus rotates relative to the fixed cosmic background could uniquely isolate the Lorentz-violating signal from isotropic BSM backgrounds.

In a broader context, although the present analysis has been formulated within relativistic quantum field theory, it illustrates a general spectral mechanism that may also be relevant in other scenarios. In particular, the sensitivity of the vacuum energy to the component of the axial background projected along the confinement direction reflects the role played by anisotropic fermionic couplings in shaping the mode structure of confined systems. Similar effects may therefore arise in systems whose low-energy excitations are described by effective Weyl or Dirac Hamiltonians, where boundary conditions and finite-size quantization play a central role in determining the spectrum.

Several extensions of the present work would be worth exploring. These include the study of finite-temperature corrections, the implementation of more general boundary conditions, and the analysis of alternative geometries where the anisotropy induced by the axial background could lead to qualitatively different vacuum responses. It would also be interesting to investigate the interplay between axial background couplings and external electromagnetic fields, as well as extensions to effective models more directly connected with topological semimetals. We hope that the results presented here contribute to clarifying the role of axial Lorentz-violating couplings in confined fermionic systems and provide a useful conceptual link between Casimir physics, effective field theory, and topological phases of matter.

%------------ Acknowledges -----------------
\acknowledgements{A.M.-R. acknowledges financial support by UNAM-PAPIIT project No. IG100224, UNAM-PAPIME project No. PE109226, by SECIHTI project No. CBF-2025-I-1862 and by the Marcos Moshinsky Foundation. M. B. Cruz acknowledges financial support from the Conselho Nacional de Desenvolvimento Cient\'{i}fico e Tecnol\'{o}gico (CNPq), through grant 301812/2026-8. E.R.B.M. thanks CNPq for partial support, Grant No. 304332/2024-0.} \\
\\
\\
\noindent\\
{\bf Declaration} The authors declare that they have no competing financial interests or personal relationships that could have influenced the work reported in this paper. \\
\\
\\
\noindent\\
{\bf Conflict of interest } The authors declare no conflict of interest

%---------- References ------------------


\begin{thebibliography}{100}

\bibitem{Casimir1948}
H.~B.~G. Casimir, On the Attraction Between Two Perfectly Conducting Plates, \textit{Proc. Kon. Ned. Akad. Wet.}, v.~51, p. 793--795, 1948.

\bibitem{Sparnaay1958}
M.~J. Sparnaay, Measurements of attractive forces between flat plates, \textit{Physica}, v.~24, p. 751--764, 1958.

\bibitem{Lamoreaux:1996wh}
S.~K. Lamoreaux, Demonstration of the Casimir Force in the 0.6 to $6\mu\text{m}$ Range, \textit{Phys. Rev. Lett.}, v.~78, p. 5--8, 1997.

\bibitem{Mohideen:1998iz}
U.~Mohideen e A.~Roy, Precision Measurement of the Casimir Force from 0.1 to $0.9\mu\text{m}$, \textit{Phys. Rev. Lett.}, v.~81, p. 4549--4552, 1998.


\bibitem{Bressi:2002fr}
G.~Bressi, G.~Carugno, R.~Onofrio, e G.~Ruoso, Measurement of the Casimir Force between Parallel Metallic Surfaces, \textit{Phys. Rev. Lett.}, v.~88, p. 041804, 2002.

\bibitem{Decca:2003zk}
R.~S. Decca, D.~L{\'o}pez, E.~Fischbach, G.~L. Klimchitskaya, D.~E. Krause, e V.~M. Mostepanenko, Precise comparison of theory and new experiment for the Casimir force leads to stronger constraints on thermal quantum effects and long-range interactions, \textit{Annals of Physics}, v.~318, n.~1, p. 37--80, 2005.

\bibitem{Milton2001}
K.~A. Milton, \textit{The Casimir Effect: Physical Manifestations of Zero-Point Energy}. Singapore: World Scientific, 2001.

\bibitem{Bordag2009}
M.~Bordag, G.~Klimchitskaya, U.~Mohideen, e V.~Mostepanenko, \textit{Advances in the Casimir Effect}. Oxford: Oxford University Press, 2009.


\bibitem{Plunien:1986ca}
G.~Plunien, B.~M{\"u}ller, e W.~Greiner, The Casimir effect, \textit{Physics Reports}, v.~134, n.~2, p. 87--193, 1986.

\bibitem{Klimchitskaya:2009zz}
G.~L. Klimchitskaya, U.~Mohideen, e V.~M. Mostepanenko, The Casimir Force between Real Materials: Experiment and Theory, \textit{Rev. Mod. Phys.}, v.~81, p. 1827--1885, 2009.

\bibitem{Birrell:1982ix}
N.~D. Birrell e P.~C.~W. Davies, \textit{Quantum Fields in Curved Space}. Cambridge: Cambridge University Press, 1982.

\bibitem{Colladay1997}
D.~Colladay e V.~A. Kosteleck{\'y}, CPT violation and the standard model, \textit{Phys. Rev. D}, v.~55, p. 6760--6774, 1997.

\bibitem{Colladay1998}
D.~Colladay e V.~A. Kosteleck{\'y}, Lorentz-violating extension of the standard model, \textit{Phys. Rev. D}, v.~58, p. 116002, 1998.

\bibitem{Kostelecky2004}
V.~A. Kosteleck{\'y}, Gravity, Lorentz violation, and the standard model, \textit{Phys. Rev. D}, v.~69, p. 105009, 2004.


\bibitem{Kostelecky:1988zi}
V.~A. Kosteleck{\'y} e S.~Samuel, Spontaneous breaking of Lorentz symmetry in string theory, \textit{Phys. Rev. D}, v.~39, p. 683--685, 1989.

\bibitem{Jacobson:2000xp}
T.~Jacobson e D.~Mattingly, Gravity with a dynamical preferred frame, \textit{Phys. Rev. D}, v.~64, p. 024028, 2001.


\bibitem{Mattingly:2005re}
D.~Mattingly, Modern Tests of Lorentz Invariance, \textit{Living Reviews in Relativity}, v.~8, n.~1, p. 5, 2005.

\bibitem{Liberati:2013xla}
S.~Liberati, Tests of Lorentz invariance: a 2013 update, \textit{Class. Quant. Grav.}, v.~30, p. 133001, 2013.

\bibitem{PhysRevD.96.045019}
M.~B. Cruz, E.~R.~Bezerra de Mello, e A.~Yu. Petrov, Casimir effects in Lorentz-violating scalar field theory, \textit{Phys. Rev. D}, v.~96, p. 045019, 2017.

\bibitem{Erdas:2020ilo}
A.~Erdas, Casimir effect of a Lorentz-violating scalar in magnetic field, 
\textit{Int. J. Mod. Phys. A} \textbf{35}, no.31, 2050209 (2020)


\bibitem{Escobar2020}
C.~A. Escobar, L.~Medel, e A.~Mart{\'i}n-Ruiz, Casimir effect in Lorentz-violating scalar field theory: A local approach, \textit{Phys. Rev. D}, v.~101, p. 095011, 2020.

\bibitem{ESCOBAR2020135567}
C.~A. Escobar, A.~M. Mart{\'i}n-Ruiz, O.~J. Franca, e M.~A.~G. Garcia, A non-perturbative approach to the scalar Casimir effect with Lorentz symmetry violation, \textit{Physics Letters B}, v.~807, p. 135567, 2020.

\bibitem{Frank2008}
M.~Frank e I.~Turan, Casimir force in a Lorentz-violating theory, \textit{Phys. Rev. D}, v.~74, p. 033016, 2006.


\bibitem{10.1093/ptep/ptae016}
A.~Rohim, A.~Romadani, e A.~S. Adam, Casimir Effect of Lorentz-Violating Charged Dirac Field in Background Magnetic Field, \textit{Progress of Theoretical and Experimental Physics}, v.~2024, p. 033B01, 2024.

\bibitem{PhysRevD.99.085012}
M.~B. Cruz, E.~R.~B. de Mello, e A.~Yu. Petrov, Fermionic Casimir effect in a field theory model with Lorentz symmetry violation, \textit{Phys. Rev. D}, v.~99, p. 085012, 2019.

\bibitem{Kharlanov2009}
O.~G. Kharlanov e V.~Ch. Zhukovsky, Casimir effect in Lorentz-violating electrodynamics, \textit{Phys. Rev. D}, v.~81, p. 025015, 2010.

\bibitem{PhysRevD.94.076010}
A.~Mart{\'i}n-Ruiz e C.~A. Escobar, Casimir effect between ponderable media as modeled by the standard model extension, \textit{Phys. Rev. D}, v.~94, p. 076010, 2016.

%\bibitem{MartinRuiz2016}
%A.~Mart{\'i}n-Ruiz e C.~A. Escobar, Local effects of the quantum vacuum in Lorentz-violating electrodynamics, \textit{Phys. Rev. D}, v.~94, p. 105019, 2016.

\bibitem{Tang:2025eew}
Y.~Tang, W.~Liu, e J.~Wang, Observational signature of Lorentz violation in acceleration radiation, \textit{Eur. Phys. J. C}, v.~85, n.~10, p. 1108, 2025.

\bibitem{Dudal2018}
D.~Dudal, A.~J. Mizher, e P.~Pais, Remarks on the Chern-Simons photon term in the QED description of graphene, \textit{Phys. Rev. D}, v.~98, n.~6, p. 065008, 2018.

\bibitem{Dudal:2020yah}
D.~Dudal, P.~Pais, e L.~Rosa, Casimir energy in terms of boundary quantum field theory: The QED case, \textit{Phys. Rev. D}, v.~102, n.~1, p. 016026, 2020.

\bibitem{Dudal2024}
D.~Dudal, A.~Gobeyn, T.~Oosthuyse, S.~Stouten, e D.~Vercauteren, Casimir energy with perfect electromagnetic boundary conditions and duality: A field-theoretic approach, \textit{Phys. Rev. D}, v.~110, n.~6, p. 065015, 2024.

\bibitem{Armitage2018}
N.~P. Armitage, E.~J. Mele, e A.~Vishwanath, Weyl and Dirac semimetals in three-dimensional solids, \textit{Rev. Mod. Phys.}, v.~90, p. 015001, 2018.

\bibitem{Grushin2012}
A.~G. Grushin, Consequences of a condensed matter realization of Lorentz-violating QED, \textit{Phys. Rev. D}, v.~86, p. 045001, 2012.

\bibitem{Goswami2013}
P.~Goswami e S.~Tewari, Axionic field theory of Weyl semimetals, \textit{Phys. Rev. B}, v.~88, p. 245107, 2013.

\bibitem{Hasan2010}
M.~Z. Hasan e C.~L. Kane, Colloquium: Topological insulators, \textit{Rev. Mod. Phys.}, v.~82, p. 3045--3067, 2010.

\bibitem{Qi2011}
X.-L. Qi e S.-C. Zhang, Topological insulators and superconductors, \textit{Rev. Mod. Phys.}, v.~83, p. 1057--1110, 2011.

\bibitem{PhysRevResearch.4.023106}
V.~A. Kosteleck{\'y}, R.~Lehnert, N.~McGinnis, M.~Schreck, e B.~Seradjeh, Lorentz violation in Dirac and Weyl semimetals, \textit{Phys. Rev. Res.}, v.~4, p. 023106, 2022.

\bibitem{PhysRevD.109.065005}
A.~G{\'o}mez, R.~Mart{\'i}nez von Dossow, A.~Mart{\'i}n-Ruiz, e L.~F. Urrutia, Lorentz invariance violation and the $CPT$-odd electromagnetic response of a tilted anisotropic Weyl semimetal, \textit{Phys. Rev. D}, v.~109, p. 065005, 2024.

\bibitem{sym17040581}
R.~Mart{\'i}nez von Dossow, A.~Mart{\'i}n-Ruiz, e L.~F. Urrutia, Higher-Order Derivative Corrections to Axion Electrodynamics in 3D Topological Insulators, \textit{Symmetry}, v.~17, n.~4, p. 581, 2025.

\bibitem{ElizaldeKirsten}
E.~Elizalde e K.~Kirsten, Casimir energy for a massive fermionic field with MIT boundary conditions, \textit{J. Math. Phys.}, v.~35, p. 1260, 1994.

\bibitem{Capasso:2007zz}
F.~Capasso, J.~N. Munday, D.~Iannuzzi, e H.~B. Chan, Casimir Forces and Quantum Electrodynamical Torques: Physics and Nanomechanics, \textit{IEEE Journal of Selected Topics in Quantum Electronics}, v.~13, n.~2, p. 400--414, 2007.

\bibitem{Woods:2016vwu}
L.~M. Woods, D.~A.~R. Dalvit, A.~Tkatchenko, P.~Rodriguez-Lopez, A.~W. Rodriguez, e R.~Podgornik, Materials perspective on Casimir and van der Waals interactions, \textit{Rev. Mod. Phys.}, v.~88, p. 045003, 2016.

\bibitem{Muniz_2025}
C.~R. Muniz, M.~B. Cruz, R.~M.~P. Neves, M.~Farooq, e M.~Zubair, Hot Casimir wormholes in Einstein Gauss-Bonnet gravity, \textit{Journal of Cosmology and Astroparticle Physics}, v.~2025, p. 015, 2025.

\bibitem{MiltonBag}
K.~A. Milton, Casimir energies and pressures for delta-function potentials, \textit{J. Phys. A}, v.~37, p. 6391, 2004.

\bibitem{BordagBag}
M.~Bordag, K.~Kirsten, e D.~V. Vassilevich, On the ground state energy for a penetrable sphere and for a dielectric ball, \textit{Phys. Rev. D}, v.~59, p. 085011, 1999.

\bibitem{PhysRevResearch.3.023201}
T.~Ishikawa, K.~Nakayama, e K.~Suzuki, Lattice-fermionic Casimir effect and topological insulators, \textit{Phys. Rev. Res.}, v.~3, p. 023201, 2021.


\end{thebibliography}
\end{document}